
\magnification=\magstep1
\hfuzz=6pt
\nopagenumbers
\font\tif=cmr10 scaled \magstep3

\rightline{PUPT-1562$|$hep-th/9509084}
\vfil
\centerline{\tif Infrared regularization of non-Abelian gauge theories}
\vfil
\centerline{{\rm Vipul
Periwal}\footnote{${}^*$}{vipul@puhep1.princeton.edu}}
\bigskip
\centerline{Department of Physics}\centerline{Princeton University
}\centerline{Princeton, New Jersey 08544-0708}
\vfil
\par\noindent
Curci and Ferrari found
a unique BRS-invariant action for non-Abelian
gauge theories which includes a mass term for the gauge bosons.  I
analyze this action.  While
the BRS operator is not nilpotent, the Zinn-Justin equation generalizes
in a simple way so that the renormalization of the theory is consistent
with the infrared regularization provided by the mass---infrared singularities
and ultraviolet infinities are therefore clearly separated.  Relations
between renormalization constants are derived in dimensional regularization
with minimal subtraction.  Additional
new symmetries  allow a simple characterization of physical operators.
A new formula is given for the gauge parameter dependence of physical
operators.

\vfil\eject
\def\half{\hbox{${1\over 2}$}}

\def\part{\partial}

\def\abc{\{b,c\}}
\def\cbc{[b,c]}

\def\lam{\lambda}
\def\pA{\partial\cdot A}

\def\tr{{\rm tr}}
\def\D{{\rm D}}

\def\ghost{3}
\def\hist{1}
\def\st{4}
\def\zjtwo{7}
\def\zjthr{8}
\def\v{9}
\def\brs{5}
\def\zj{6}
\def\text{2}
\def\newref{10}
\def\newreff{11}
\def\curci{12}
The infrared behaviour of non-Abelian gauge theories, in the absence of
masses generated by spontaneous symmetry breaking in matter sectors, has
been a puzzle for more than two decades.  It is believed that
gluons are `confined', {\it i.e.} coloured particles do not appear
in the observable spectrum
of particles, but there is nothing resembling a proof that confinement
occurs in continuum non-Abelian gauge theories.  The phenomenon of
confinement is believed to be linked to the growth of the coupling
parameter as the normalization point is taken to longer length scales,
an intuition derived from weak-coupling perturbation theory.
The severe on-shell divergences of amplitudes with external gluons are
also viewed as hints of confinement, but again, this has not been translated
into a rational argument for confinement.
A simple regularization of the infrared divergences, independent
of the ultraviolet problems, would illuminate the physics of such
on-shell divergences.

In the present paper, I address the problem of quantitatively
regularizing the infrared divergences of gauge theories in a manner
compatible with the renormalization of ultraviolet divergences.
The action I use was found by Curci and Ferrari[\newref].  They
showed that mass terms for gluons can be incorporated into a {\it unique}
BRS-invariant[\brs] action.   They also discussed the renormalizability
of the theory.  Indeed, as will be reviewed here, the only aspect of the
regularization and renormalization of the theory that is changed is the
addition of an inhomogeneous term in the Zinn-Justin[\zjtwo] equation.
The fact that the Slavnov-Taylor[\st] identities are simple in a particular
gauge for a particular form of symmetry breaking (the mass term)
will not come as a surprise.   Using the massive Zinn-Justin equation,
I show that the theory can be multiplicatively renormalized.
(For a clear explanation of the effects of mass terms in
the quantization of non-Abelian gauge bosons, see Ref.\hist,\newreff.)

Some of the
analysis I present is already contained in Ref.~[\newref].  The new aspects
in the present paper are: (1) The U(1) symmetry, and its applications;
(2) the formula for the dependence of physical observables on $\xi;$
and (3) the massive Zinn-Justin equation, and
the relations between renormalization
constants in dimensional regularization with minimal subtraction.
I am grateful to M. Schaden for informing me of the existence of Ref.~\newref,
\newreff.

The organization of this paper is as follows: I  first exhibit an
action and a set of BRS variations with the advertized properties.  I
then point out several interesting features of this action.  I derive
the appropriate form of the Zinn-Justin equation for this action.  Finally,
I solve the Zinn-Justin equation to determine the form of the
divergent part of the effective action, thus determining relations between
the renormalization constants.

It is a textbook[\text] fact that mass terms for photons are compatible with
BRS invariance in Lorentz gauge.  BRS variations for photons are
$$sA_m = \part_mc,\qquad sc= 0,\qquad sb={\cal F},$$
where $c$ is the anticommuting ghost field, $b$ is the anticommuting antighost
field, and ${\cal F}$ is the gauge condition.  Then
$$s\half\int A^2 +[b,c]=\int A_m\part^mc +{\cal F}c=\int(-\part\cdot A
+{\cal F})c.$$
Thus, in Lorentz gauge a mass term for the photon
is compatible with BRS invariance, since the rest of the gauge-fixed action
is obviously invariant.  Not all is well though, since now
$$s^2b = s\part\cdot A =\part^2c\not={{\delta S}\over{\delta b}},$$
since $S$ now contains a mass term.

What is the analogue of this gauge choice for
non-Abelian gauge theories?
The BRS transformations are as follows:
$$sA_m = D_mc=\part_mc+[A_m,c],\qquad sc = -c^2, \qquad sb ={\cal F},$$
with all fields in the adjoint representation.
This implies
$$s\int \half \tr \left(A^2 +\cbc\right)
= \int \tr \left[ c\left(-\pA +\half \abc + {\cal F}\right)\right].$$
Therefore, a mass for a gauge boson would be compatible with BRS invariance in
the gauge specified by
${\cal F}= \pA -\half \abc.$  This is a specific case of the general
nonlinear gauge choices studied by Zinn-Justin[\zj].
\def\fr#1#2{\hbox{${{#1}\over{#2}}$}}

Given a nonlinear gauge of this form, {\it i.e.} for a choice of the
BRS transformation of the antighost, $b,$ which involves $b,$
it is not a priori obvious that there is
a BRS invariant action.  In the case at hand, in fact there is such an
action,
$$S= S_{ym} + S_m
+ {1\over 2g^2\xi}\int \tr \left[b\part^2c + \half j_mA^m + \fr18 \abc^2 -
\half (\part\cdot A)^2\right] .$$
Here $S_m\equiv -(m^2/4g^2)\int \tr\left(A^2 +\cbc\right),$
$j^m \equiv \{c,\part_mb\} -\{b,\part_mc\}$ is the current associated with
global rotations of the ghost-antighost fields, and
$S_{ym}$ is the classical Yang-Mills action.
The adjoint representation has been normalized so that $\tr T_iT_j =
-2\delta_{ij}.$

\def\bs{{\bar s}}
\def\hs{{\hat s}}
\def\eps{\epsilon}
Some features of $S_{gf} \equiv S-S_{ym}-S_m$ are worth noting:
\item{1.} There is a U(1) symmetry, that leaves $A$ invariant, with
$$\delta b = \eps c ,\qquad \delta c = -\eps b.$$
The $\pi/2$ rotation is the obvious
discrete symmetry $b\rightarrow c, c\rightarrow -b.$
Due to this symmetry, there is an additional fermionic
invariance of $S,$
$$\bs A_m = -D_mb,\qquad \bs b = b^2,\qquad \bs c=\part\cdot A + \half\abc.$$
$s$ and $\bs$ do not anticommute.
\item{2.} Define observables to be
those operators of vanishing ghost number
which are invariant under this U(1) symmetry, and are annihilated by $s.$
Such observables are also annihilated by $\bs.$
It is an immediate consequence that correlation functions
of physical observables ${\cal O}_\alpha:
s{\cal O}_\alpha=\bs{\cal O}_\alpha=0 $ with $\pA$ vanish:
$$\bigg\langle \pA \prod_\alpha{\cal O}_\alpha\bigg\rangle = 0.$$
It can be shown that observables satisfy
${\cal O}_\alpha={\cal O}_\alpha(A).$
\item{3.} $S_{gf}$ can be written as a sum in the form
$$S_{gf}= {1\over 2g^2\xi}
\int \tr\big[ s(-b\part\cdot A) + \half (\part\cdot A)^2 + \fr18\abc^2 \big].$$
\item{4.} By introducing a Lagrange multiplier, and a modified BRS
operator $\hs$ defined by $\hs A=sA,$ $\hs c=sc,$ $\hs b=\lam,$ and
$\hs\lam=0,$ it is possible to write
$$S_{gf} = {1\over 2g^2\xi}\int\tr\left(
\hs \left[b(\half\lam-\pA+\fr14\abc)\right]
-\half(\lam-\pA+\half\abc)^2\right).$$
Now
$\hs S_m = -(1/2g^2)\int\tr\left[c\left(\lam-\pA+\half\abc\right)
\right].$  Even with the Lagrange multiplier the action has an additional
fermionic symmetry, derived from $\bs.$  This is also true if one
adds an additional auxiliary field to eliminate quartic ghost-antighost
interactions, following Zinn-Justin[\zj].
\item{6.} It can be shown that correlation functions of physical observables
depend on $\xi$ as
$$ \xi{\part\over\part\xi}\left\langle \prod_\alpha{\cal O}_\alpha
\right\rangle
= {m^2\over 8g^2}\left\langle\int\tr\cbc
\prod_\alpha{\cal O}_\alpha\right\rangle.$$
This equation is a relation between {\it bare} quantities.
Thus the expected $\xi$ independence is recovered when $m^2=0.$
Compare this result to the analysis in Curci and Ferrari[\newref].

\def\dede#1#2{{{\delta{#1}}\over{\delta{#2}}}}
\def\cS{{\cal S}}
\def\cF{{\cal F}}
I now derive the Zinn-Justin(ZJ) equation for this action.  Recall that
this is the simplest formulation of the regularization and renormalization
of non-Abelian gauge theories[\zjthr].
It is convenient to rescale fields to absorb
powers of $g$ and $\xi,$ so we redefine $A\rightarrow gA, b\rightarrow g\xi b,$
and $c\rightarrow gc.$  Now $s$ is given by
$$sA_m = D_mc = \part_mc +g[A_m,c],\qquad sc = -gc^2,\qquad sb = {1\over\xi}
\left(\part\cdot A - \half g\xi\abc\right)\equiv \cF.$$
The ZJ equation is a quadratic equation satisfied by
the generating functional of proper vertices.  It encapsulates in an elegant
manner the content of the Slavnov-Taylor identities.
To derive the ZJ equation, I introduce sources for the fields, $A,b,c,$ and
for their BRS variations, so let
$S_J \equiv \int \tr\big[ J_A^m A_m + b J_b + J_c c\big],$
$\cS \equiv S + \int \tr \big[K_A^m D_mc + K_b \cF -K_c gc^2\big],$ with
$$sS_J = \int\tr \bigg[J_A^m \dede \cS{K_A^m}+
\dede \cS{K_b} J_b -J_c \dede\cS{K_c}\bigg],\
\quad s\cS = \int\tr \big[K_b s\cF\big] =
{2\over\xi}\int\tr K_b\bigg[\dede \cS{b} +
{ m^2\xi\over2}c\bigg].$$
I use dimensional regularization to define the functional integral
$$\exp(W) \equiv \int \D A\D b\D c \exp\left(-\cS +S_J\right).$$

\def\tS{\tilde {\cal S}}
It follows from the invariance of the regularized measure that
$\big\langle s\big(S_J - \cS\big)\exp(S_J)\big\rangle=0.$
Therefore, defining
$$M\ast N \equiv \int\tr\bigg(\dede M{K_A^m}\dede N{A_{m}}+
\dede M{K_b}\dede N{b}+\dede M{K_c}\dede Nc\bigg),$$
I find
$$\left\langle \left(\cS\ast\cS -\int {2K_b\over\xi}\bigg[\dede \cS{b} +
{ m^2\xi\over2}c\bigg]\right)\exp(S_J)\right\rangle = 0.$$
It is straightforward to
verify that the action $\cS$ obeys this equation {\it independent} of the
functional integral.
Before proceeding with the argument for renormalizability, it is convenient
to define $\tS \equiv \cS- \int\tr K_b^2/\xi,$ which satisfies
$\tS\ast\tS= m^2\int\tr K_bc.$

\def\al{\alpha}
\def\zz#1{Z^{#1}_\ell}
\def\zo#1{Z^{#1}_{\ell-1}}
\def\di{{\rm div}}
\def\tG{\tilde\Gamma}
{}From the regularized functional integral form, and using
the definition of the generating functional of proper vertices,
$$\Gamma[A,b,c,K_A,K_b,K_c]+W[J_A,J_b,J_c,K_A,K_b,K_c]= \int \tr\left[
J_AA + bJ_b+J_cc\right],$$
I  find that $\tG \equiv \Gamma- \int\tr K_b^2/\xi$ satisfies
$\tG\ast\tG= m^2\int\tr K_bc.$
I now go through the standard induction
argument[\zjthr] for
the renormalization of this massive version of non-Abelian gauge theory.
As the starting point of the induction,
$\Gamma_0$ and $\cS$ satisfy the massive Zinn-Justin equation, and
$\Gamma_0=\cS$ is finite.
So supposing that $\tG$ has been rendered finite to $\ell-1$ loop
order, the divergent part of the $\ell$ loop effective action satisfies
$$\tS\ast\tG_{\ell}^{\di} + \tG_{\ell}^{\di}\ast\tS=0.$$
This equation is solved by writing down all possible terms
restricted by dimensional analysis, global symmetry, and
ghost number conservation.
The counterterms required to render $\tG$ finite to $\ell$ loop order
are such that
$$\tS_\ell = \tS_{\ell-1} - \tG_{\ell}^{\di} + \hbox{higher orders}.$$
The higher order terms are required in order that $\tS_\ell$
satisfies $\tS_\ell\ast\tS_\ell=m^2\int\tr K_b c,$ since we must
ensure that $\tG\ast\tG=m^2\int\tr K_b c$ in order for the induction to
proceed.  By solving the ZJ equation, I find that $\tG_{\ell}^{\di}$ is such
that
$$\tS_\ell[A,b,c,K_A,K_b,K_c;g,\xi,m^2] =
\tS[AZ^{A}_\ell,b\zz b,c\zz b,K_A\zz \kappa,K_b\zz K,
K_c\zz K; g\zz g,\xi\zz \xi , m^2\zz m],$$
with
$$\eqalign{\zz A &= \zo A -\alpha_A ,\cr
\zz g &= \zo g -\al_g,\cr
\zz b &= \zo b - \al_b,\cr
\zz \xi &= \zo \xi +6\al_b+\al_A+3\al_g,\cr
\zz m &=\zo m -6\al_b-2\al_A-4\al_g,\cr
\zz \kappa &=\zo \kappa +3\al_b+2\al_A+2\al_g,\cr
\zz K &=\zo K + 4\al_b +\al_A+2\al_g.\cr}$$
Note that the nonlinear gauge has resulted in strong relations between
renormalization constants.  There are just three independent renormalization
constants.  The proof is complete, since I
have rendered $\tG_\ell$ finite, with counterterms that preserve the
induction hypothesis.

In conclusion, I have studied the unique action with
BRS invariance and a mass term for gluons[\newref].  I found
additional
symmetries that may be used to characterize physical observables, which
are found to be functions of the gauge potential alone.
I have shown that the
theory can be renormalized in the usual manner, deriving relations between
renormalization constants in dimensional regularization with minimal
subtraction.  I derived a simple formula for the
$\xi$ dependence of physical observables, showing that
$\xi$ independence is restored in the limit $m^2\downarrow
0.$  Unitarity is a delicate question in the massless theory;
for more on unitarity,  see Ref.~\newref,\newreff.  For a different
approach to unitarity, see Ref.~\v.

The physical application is to the analysis of infrared
divergences in non-Abelian gauge theories with the use of this
action, as in Ref.~\curci.  One might also
consider
infrared renormalons and the measure for instantons with this gluon mass.
Another application to particle physics may be found in Ref.~\v.

\bigskip
I thank R. Myers for collaborating in the early stages of this work,
and B. Ratra for helping me check some computations.
This work was supported in part by NSF Grant No. PHY90-21984.

\bigskip
\centerline{References}
\medskip

\item{\hist} A.A. Slavnov and L.D. Faddeev, {\sl Teor. Mat. Fiz.}
{\bf 3} (1970) 18

\item{\text} P. Ramond, {\it Field theory: a modern primer},
Benjamin/Cummings, (Reading, MA,  1981)

\item{\ghost} L.D. Faddeev and V.N. Popov, {\sl Phys. Lett.} {\bf 25B}
(1967) 29; V.N. Popov and L.D. Faddeev,
Kiev Report ITP-67-36 (1967), (English transl. NAL-THY-57
(1972), tr. D. Gordon and B.W. Lee)

\item{\st} A.A. Slavnov, {\sl Theor. Math. Phys.} {\bf 10} (1972) 99;
J.C. Taylor, {\sl Nucl. Phys.} {\bf B33} (1971) 436

\item{\brs} C. Becchi, A. Rouet and R. Stora, {\sl Ann. Phys.} {\bf 98}
(1976) 287

\item{\zj} J. Zinn-Justin, {\sl Nucl. Phys.} {\bf B246} (1984) 246

\item{\zjtwo} J. Zinn-Justin,  in {\it Trends in Elementrary Particle Theory},
H. Rollnick and K. Dietz (eds.), Springer Verlag (Berlin, 1975)

\item{\zjthr} J. Zinn-Justin, {\it Quantum Field Theory and Critical
Phenomena}, Oxford University Press (Oxford, 1993)

\item{\v} V. Periwal, {\it Unitary theory of massive non-Abelian vector
bosons}, Princeton preprint PUPT-1563 (1995)

\item{\newref} G. Curci and R. Ferrari, {\sl Nuo. Cim.} {\bf 32A} (1976) 151;
{\bf 32A} (1976) 1

\item{\newreff} I. Ojima, {\sl Z. Phys.} {\bf C13} (1982) 173;
for a good review, see R. Delbourgo, S. Twisk, G. Thompson,
{\sl Int. J. Mod. Phys.} {\bf A3} (1988) 435, and other references therein

\item{\curci} G. Curci and E. d'Emilio, {\sl Phys. Lett.} {\bf 83B} (1979) 199
\end